\documentclass[12pt]{article}
\usepackage{amssymb}
\usepackage{amsfonts}
\usepackage{amsmath}

\begin{document}

\title{The fractional volatility model: No-arbitrage, leverage and
completeness}
\author{R. Vilela Mendes\thanks{%
Centro de Matem\'{a}tica e Aplica\c{c}\~{o}es Fundamentais, Universidade de
Lisboa, Av. Prof. Gama Pinto 2, P1649-003 Lisboa} \thanks{%
IPFN, EURATOM/IST Association, e-mail: rvilela.mendes@gmail.com}, M. J. Oliveira%
\footnotemark[1] \thanks{%
Universidade Aberta, oliveira@cii.fc.ul.pt} and A.M. Rodrigues%
\footnotemark[1]}
\date{}
\maketitle

\begin{abstract}
Based on a criterion of mathematical simplicity and consistency with
empirical market data, a stochastic volatility model has been obtained with
the volatility process driven by fractional noise. Depending on whether the
stochasticity generators of log-price and volatility are independent or are
the same, two versions of the model are obtained with different leverage
behavior. Here, the no-arbitrage and completeness properties of the models
are studied.
\end{abstract}


\section{Introduction}

In liquid markets the autocorrelation of price changes decays to negligible
values in a few minutes, consistent with the absence of long term
statistical arbitrage. Because innovations of a martingale are uncorrelated,
there is a strong suggestion that it is a process of this type that should
be used to model the stochastic part of the returns process. As a
consequence, classical Mathematical Finance has, for a long time, been based
on the assumption that the price process of market securities may be
approximated by geometric Brownian motion 
\begin{equation}
\begin{array}{lll}
dS_{t} & = & \mu S_{t}dt+\sigma S_{t}dB\left( t\right)%
\end{array}
\label{1.1}
\end{equation}%
Geometric Brownian motion (GBM) models the absence of linear correlations,
but otherwise has some serious shortcomings. It does not reproduce the
empirical leptokurtosis nor does it explain why nonlinear functions of the
returns exhibit significant positive autocorrelation. For example, there is
volatility clustering, with large returns expected to be followed by large
returns and small returns by small returns (of either sign). This, together
with the fact that autocorrelations of volatility measures decline very
slowly \cite{Ding2}, \cite{Harvey}, \cite{Crato} has the clear implication
that long memory effects should somehow be represented in the process and
this is not included in the geometric Brownian motion hypothesis. The
existence of an essential memory component is also clear from the failure of
reconstruction of a Gibbs measure and the need to use chains with complete
connections in the phenomenological reconstruction of the market process 
\cite{Vilela1}.

As pointed out by Engle \cite{Engle}, when the future is uncertain investors
are less likely to invest. Therefore uncertainty (volatility) would have to
be changing over time. The conclusion is that a dynamical model for
volatility is needed and $\sigma $ in Eq.(\ref{1.1}), rather than being a
constant, becomes itself a process. This idea led to many deterministic and
stochastic models for the volatility (\cite{Taylor}, \cite{Engle2} and
references therein).

The stochastic volatility models that were proposed described some partial
features of the market data. For example leptokurtosis is easy to fit but
the long memory effects are much harder. On the other hand, and in contrast
with GBM, some of the phenomenological fittings of historical volatility
lack the kind of nice mathematical properties needed to develop the tools of
mathematical finance. In an attempt to obtain a model that is both
consistent with the data and mathematically sound, a new approach was
developed in \cite{Oliveira}. Starting only with some criteria of
mathematical simplicity, the basic idea was to let the data itself tell us
what the processes should be.

The basic hypothesis for the model construction were:

(H1) The log-price process $\log S_{t}$ belongs to a probability product
space $(\Omega _{1}\times \Omega _{2},P_{1}\times P_{2})$ of which the $%
(\Omega _{1},P_{1})$ is the Wiener space and the second one, $(\Omega
_{2},P_{2})$, is a probability space to be reconstructed from the data.
Denote by $\omega _{1}\in \Omega _{1}$ and $\omega _{2}\in \Omega _{2}$ the
elements (sample paths) in $\Omega _{1}$ and $\Omega _{2}$ and by $\mathcal{F%
}_{1,t}$ and $\mathcal{F}_{2,t}$ the $\sigma $-algebras in $\Omega _{1}$ and 
$\Omega _{2}$ generated by the processes up to $t$. Then, a particular
realization of the log-price process is denoted 
\begin{equation*}
\log S_{t}\left( \omega _{1},\omega _{2}\right)
\end{equation*}%
This first hypothesis is really not limitative. Even if none of the
non-trivial stochastic features of the log-price were to be captured by
Brownian motion, that would simply mean that $S_{t}$ was a trivial function
in $\Omega _{1}$.

(H2) The second hypothesis is stronger, although natural. It is assumed that
for each fixed $\omega_2$, $\log S_{t}\left( \cdot ,\omega_2\right) $ is a
square integrable random variable in $\Omega_1$.

These principles and a careful analysis of the market data led, in an
essentially unique way\footnote{%
Essentially unique in the sense that the empiricaly reconstructed volatility
process is the simplest one, consistent with the scaling properties of the
data.}, to the following model:%
\begin{equation}
dS_{t}=\mu _{t}S_{t}\,dt+\sigma _{t}S_{t}\,dB\left( t\right)  \label{1.2a}
\end{equation}%
\begin{equation}
\log \sigma _{t}=\beta +\frac{k}{\delta }\left\{ B_{H}\left( t\right)
-B_{H}\left( t-\delta \right) \right\}  \label{1.2b}
\end{equation}

the data suggesting values of $H$ in the range $0.8-0.9$. In this coupled
stochastic system, in addition to a mean value, volatility is driven by
fractional noise. Notice that this empirically based model is different from
the usual stochastic volatility models which assume the volatility to follow
an arithmetic or geometric Brownian process. Also in Comte and Renault \cite%
{Comte} and Hu \cite{Hu}, it is fractional Brownian motion that drives the
volatility, not its derivative (fractional noise). $\delta $ is the
observation scale of the process. In the $\delta \rightarrow 0$ limit the
driving process would be a distribution-valued process.

The equation (\ref{1.2b}) leads to 
\begin{equation}
\sigma _{t}=\theta e^{\frac{k}{\delta }\left\{ B_{H}\left( t\right)
-B_{H}\left( t-\delta \right) \right\} -\frac{1}{2}\left( \frac{k}{\delta }%
\right) ^{2}\delta ^{2H}}  \label{1.4}
\end{equation}%
with $E\left[ \sigma _{t}\right] =\theta >0$.

The model has been shown \cite{Oliveira} to describe well the statistics of
price returns for a large $\delta $-range and a new option pricing formula,
with "smile" deviations from Black-Scholes, was also obtained. An
agent-based interpretation \cite{Vilela2} also led to the conclusion that
the statistics generated by the model was consistent with the limit order
book price setting mechanism.

To our surprise, this data-reconstructed model was met with some hostility
by the Mathematical Finance community. Perhaps in part because much of the
nice results in this field are based on simple GBM and also because
fractional Brownian has been associated to arbitrage. In fact, in the past,
several authors tried to describe the long memory effect by replacing in the
price process Brownian motion by fractional Brownian motion with $H>1/2$.
However it was soon realized \cite{Rogers}, \cite{Shiryaev}, \cite{Salopek}, 
\cite{Sottinen} that this replacement implied the existence of arbitrage.
These results might be avoided either by restricting the class of trading
strategies \cite{Cheridito}, introducing transaction costs \cite{Guasoni} or
replacing pathwise integration by a different type of integration \cite%
{Oksendal} \cite{Elliot}. However this is not free of problems because the
Skorohod integral approach requires the use of a Wick product either on the
portfolio or on the self-financing condition, leading to unreasonable
situations from the economic point of view (for example positive portfolio
with negative Wick value, etc.) \cite{Hult}.

The fractional volatility model in Eqs.(\ref{1.2a}-\ref{1.2b}) is not
affected by these considerations, because it is the volatility process that
is driven by fractional noise, not the price process. In fact a no-arbitrage
result may be proven. This is no surprise because our requirement (H2) that,
for each sample path $\omega _{2}\in \Omega _{2}$, $\log S_{t}\left( \cdot
,\omega _{2}\right) $ is a square integrable random variable in $\Omega _{1}$
already implies that $\int \sigma _{t}dB_{t}$ is a martingale. The square
integrability is also essential to guarantee the possibility of
reconstruction of the $\sigma $ process from the data, because it implies 
\cite{Nualart}%
\begin{equation}
\begin{array}{lll}
\frac{dS_{t}}{S_{t}}\left( \cdot ,\omega _{2}\right)  & = & \mu _{t}\left(
\cdot ,\omega _{2}\right) dt+\sigma _{t}\left( \cdot ,\omega _{2}\right)
dB_{t}%
\end{array}
\label{2.2}
\end{equation}

Our aim in this paper is to give a solid mathematical construction of the
fractional volatility model, discussing existence questions, arbitrage and
market completeness.

\section{No-arbitrage and incompleteness}

Let $\left( \Omega _{1},\mathcal{F}_{1},P_{1}\right) $ be the complete
filtered Wiener probability space, carrying a standard Brownian motion $%
B=\left( B_{t}\right) _{0\leq t<\infty }$ and a filtration $\mathbb{F}%
_{1}=\left( \mathcal{F}_{1,t}\right) _{0\leq t<\infty }$. Let also $\left(
\Omega _{2},\mathcal{F}_{2},P_{2}\right) $ be another probability space
associated to a fractional Brownian motion $B_{H}$ with Hurst parameter $%
H\in \left( 0,1\right) $ and a filtration $\mathbb{F}_{2}=\left( \mathcal{F}%
_{2,t}\right) _{0\leq t<\infty }$ generated by $B_{H}$.

Let us now embed these two probability spaces in a product space $\left( 
\overline{\Omega },\overline{\mathcal{F}},\overline{P}\right) $, where $%
\overline{\Omega }$ is the Cartesian product $\Omega _{1}\times \Omega _{2}$
and $\overline{P}$ is the product measure $P_{1}\otimes P_{2}$. We also
introduce $\pi _{1}$ and $\pi _{2}$, the projections of $\overline{\Omega }$
onto $\Omega _{1}$ and $\Omega _{2}$, as well as the $\sigma -$algebra $%
\mathcal{N}$ generated by all null sets from the product $\sigma -$algebra $%
\mathcal{F}_{1}\otimes \mathcal{F}_{2}$, that is,%
\begin{equation*}
\mathcal{N}=\sigma \left( \left\{ F\subseteq \Omega _{1}\times \Omega
_{2}|\exists G\in \mathcal{F}_{1}\otimes \mathcal{F}_{2}\text{ such that }%
F\subseteq G\text{ and }\left( P_{1}\otimes P_{2}\right) \left( G\right)
=0\right\} \right) .
\end{equation*}%
Moreover, we let $\overline{\mathcal{F}}=\left( \mathcal{F}_{1}\otimes 
\mathcal{F}_{2}\right) \vee \mathcal{N}$, the $\sigma -$algebra generated by
the union of the $\sigma -$algebras $\mathcal{F}_{1}\otimes \mathcal{F}_{2}$
and $\mathcal{N}$. Then $\overline{\mathbb{F}}=\left( \overline{\mathcal{F}}%
_{t}\right) _{0\leq t<\infty }$ is the filtration for $\overline{\mathcal{F}}%
_{t}=\left( \mathcal{F}_{1,t}\otimes \mathcal{F}_{2,t}\right) \vee \mathcal{N%
}$.

Furthermore, we extend $B$ and $B_{H}$ to $\overline{\mathbb{F}}-$adapted
processes on $\left( \overline{\Omega },\overline{\mathcal{F}},\overline{P}%
\right) $ by $\overline{B}\left( \omega _{1},\omega _{2}\right) =\left(
B\circ \pi _{1}\right) \left( \omega _{1},\omega _{2}\right) $ and $%
\overline{B}_{H}\left( \omega _{1},\omega _{2}\right) =\left( B_{H}\circ \pi
_{2}\right) \left( \omega _{1},\omega _{2}\right) $ for $\left( \omega
_{1},\omega _{2}\right) \in \overline{\Omega }$. Then, it is easy to prove
that $\overline{B}$ and $\overline{B}_{H}$ are Brownian and fractional
Brownian motions with respect to $\overline{P}$ and are independent. For
notational simplicity, hereafter $B$ and $B_{H}$ will stand for $\overline{B}
$ and $\overline{B}_{H}$.

We now consider a market with a risk-free asset with dynamics given by%
\begin{equation}
dA_{t}=rA_{t}\,dt\hspace{0.5cm}A_{0}=1  \label{2.3}
\end{equation}%
with $r>0$ constant and a risky asset with price process $S_{t}$ given by
Eqs.(\ref{1.2a})-(\ref{1.2b}), with $\mu _{t}$ a $\overline{\mathbb{F}}$%
-adapted process with continuous paths, $k$ the volatility intensity
parameter and $\delta $ the observation time scale of the process.

Eq.(\ref{1.2b}) leads to Eq.(\ref{1.4})%
\begin{equation*}
\sigma _{t}=\theta e^{\frac{k}{\delta }\left\{ B_{H}\left( t\right)
-B_{H}\left( t-\delta \right) \right\} -\frac{1}{2}\left( \frac{k}{\delta }%
\right) ^{2}\delta ^{2H}}
\end{equation*}%
with $\theta =\mathbb{E}_{\overline{P}}\left[ \sigma _{t}\right] $, hence $%
\sigma _{t}$ is a measurable and an $\overline{\mathbb{F}}-$adapted process
satisfying for all $0\leq t<\infty $%
\begin{eqnarray*}
\mathbb{E}_{\overline{P}}\left[ \int_{0}^{t}\sigma _{s}^{2}ds\right] 
&=&\int_{0}^{t}\theta ^{2}e^{-\left( \frac{k}{\delta }\right) ^{2}\delta
^{2H}}\mathbb{E}_{\overline{P}}\left[ e^{\frac{2k}{\delta }\left\{
B_{H}\left( s\right) -B_{H}\left( s-\delta \right) \right\} }\right] ds \\
&=&\theta ^{2}\exp \left\{ \left( \frac{k}{\delta }\right) ^{2}\delta
^{2H}\right\} t<\infty 
\end{eqnarray*}%
using Fubini's theorem and the moment generating function of the Gaussian
random variable $B_{H}\left( s\right) -B_{H}\left( s-\delta \right) $.

Moreover $\int_{0}^{t}\left\vert \mu _{s}\right\vert ds$ being finite $%
\overline{P}-$almost surely for $0\leq t<\infty $, an application of It\^{o}%
's formula yields%
\begin{equation*}
S_{t}=S_{0}\exp \left\{ \int_{0}^{t}\left( \mu _{s}-\frac{1}{2}\sigma
_{s}^{2}\right) ds+\int_{0}^{t}\sigma _{s}dB_{s}\right\} 
\end{equation*}

\noindent \textbf{Lemma 2.1. }\textit{Consider the measurable, }$\overline{%
\mathbb{F}}$\textit{-adapted process defined by}%
\begin{equation}
\gamma _{t}=\frac{r-\mu _{t}}{\sigma _{t}},\hspace{0.5cm}0\leq t<\infty 
\label{2.4a}
\end{equation}%
\textit{with }$\mu \in L^{\infty }\left( \left[ 0,T\right] \times \overline{%
\Omega }\right) $\textit{\ and denote by }$\eta =\left( \eta _{t}\right)
_{0\leq t<\infty }$\textit{\ the stochastic exponential of }$\left(
\int_{0}^{t}\gamma _{s}dB_{s}\right) _{0\leq t<\infty }$, \textit{that is,}%
\begin{equation}
\eta _{t}=\exp \left\{ \int_{0}^{t}\gamma _{s}dB_{s}-\frac{1}{2}%
\int_{0}^{t}\gamma _{s}^{2}ds\right\} ,\hspace{0.5cm}0\leq t<\infty 
\label{2.4c}
\end{equation}%
\textit{Then,}%
\begin{equation}
\mathbb{E}_{\overline{P}}\left[ \frac{1}{2}\int_{0}^{T}\gamma _{s}^{2}ds%
\right] <\infty ,\hspace{0.5cm}0\leq T<\infty   \label{2.4b}
\end{equation}

\textit{Proof}: We make of use the fact that the sample paths of the
fractional Brownian motion $B_{H}$ are H\"{o}lder continuous of any order $%
\alpha \geq 0$ strictly less than $H$, that is, there is $C_{\alpha }>0$
such that, $\overline{P}-$almost surely, $|B_{H}\left( t\right) -B_{H}\left(
s\right) |\leq C_{\alpha }|t-s|^{\alpha }$, for every $t,s\in \left[
0,\infty \right) $. Then, for all $0\leq T<\infty $%
\begin{eqnarray*}
&&\mathbb{E}_{\overline{P}}\left[ \frac{1}{2}\int_{0}^{T}\gamma _{s}^{2}ds%
\right]  \\
&\leq &\mathbb{E}_{\overline{P}}\left[ \exp \left\{ \frac{e^{k^{2}\delta
^{2H-2}}}{2\theta ^{2}}\int_{0}^{T}\left( r+\left\vert \mu _{s}\right\vert
\right) ^{2}e^{-2\frac{k}{\delta }\left( B\left( s\right) -B\left( s-\delta
\right) \right) }ds\right\} \right]  \\
&\leq &\mathbb{E}_{\overline{P}}\left[ \exp \left\{ \frac{\left(
r+\left\Vert \mu _{s}\right\Vert _{\infty }\right) ^{2}}{2\theta ^{2}}%
e^{k^{2}\delta ^{2H-2}}\int_{0}^{T}e^{-2\frac{k}{\delta }\left( B\left(
s\right) -B\left( s-\delta \right) \right) }ds\right\} \right]  \\
&\leq &\exp \left\{ \frac{T\left( r+\left\Vert \mu _{s}\right\Vert _{\infty
}\right) ^{2}}{2\theta ^{2}}e^{k^{2}\delta ^{2H-2}-2C_{\alpha }k\delta
^{\alpha -1}}\right\} <\infty 
\end{eqnarray*}%
$\hspace{12.5cm}\blacksquare $

Additionally, we assume that investors are allowed to trade only up to some
fixed finite planning horizon $T>0$.

\noindent \textbf{Proposition 2.2.} \textit{The market defined by (\ref{1.2a}%
), (\ref{1.2b}) and (\ref{2.3}) is free of arbitrage}

\textit{Proof:} Because, by the lemma above the process $\gamma $ in (\ref%
{2.4a}) satisfies the Novikov condition (\ref{2.4b}), the nonnegative
continuous supermartingale $\eta $ in (\ref{2.4c}) is a true $\overline{P}-$%
martingale. Hence we can define for each $0\leq T<\infty $ a new probability
measure $Q_{T}$ on $\overline{\mathcal{F}}_{T}$ by%
\begin{equation}
\frac{dQ_{T}}{d\overline{P}}=\eta _{T},\hspace{0.5cm}\overline{P}-a.s.
\label{2.5}
\end{equation}%
Then, by the Cameron-Martin-Girsanov theorem, for each fixed $T\in \left[
0,\infty \right) $, the process%
\begin{equation}
B_{t}^{\ast }=B_{t}-\int_{0}^{t}\frac{r-\mu _{s}}{\sigma _{s}}\,ds\hspace{%
0.5cm}0\leq t\leq T  \label{2.6}
\end{equation}%
is a Brownian motion on the probability space $\left( \overline{\Omega },%
\overline{\mathcal{F}}_{T},Q_{T}\right) $.

Consider now the discounted price process%
\begin{equation*}
Z_{t}=\frac{S_{t}}{A_{t}}\hspace{0.5cm}0\leq t\leq T
\end{equation*}%
Under the new probability measure $Q_{T}$, equivalent to $\overline{P}$ on $%
\overline{\mathcal{F}}_{T}$, its dynamics is given by%
\begin{equation}
Z_{t}=Z_{0}+\int_{0}^{t}\sigma _{s}Z_{s}\,dB_{s}^{\ast }  \label{2.7}
\end{equation}%
and is a martingale in the probability space $\left( \overline{\Omega },%
\overline{\mathcal{F}}_{T},Q_{T}\right) $ with respect to the filtration $%
\left( \overline{\mathcal{F}}_{t}\right) _{0\leq t<T}$. By the fundamental
theorem of asset pricing, the existence of an equivalent martingale measure
for $Z_{t}$ implies that there are no arbitrages, that is, $\mathbb{E}%
_{Q_{T}}\left[ Z_{t}|\overline{\mathcal{F}}_{s}\right] =Z_{s}$ for $0\leq
s<t\leq T$.\hfill $\blacksquare $

Another important concept is market completeness. We note that, in this
financial model, trading takes place only in the stock and in the money
market and, as a consequence, volatility risk cannot be hedged. Hence, since
there are more sources of risk than tradable assets, in this model, the
market is incomplete, as proved in the next proposition.

\noindent \textbf{Proposition 2.3.} \textit{The market defined by (\ref{1.2a}%
),(\ref{1.2b}) and (\ref{2.3}) is incomplete}

\textit{Proof}\textbf{: }Here we use an integral representation for the
fractional Brownian motion \cite{Decreusefond}, \cite{Embrechts} 
\begin{equation}
B_{H}\left( t\right) =\int_{0}^{t}K_{H}\left( t,s\right) dW_{s}
\label{2.10a}
\end{equation}%
$W_{t}$ being a Brownian motion with respect to $\overline{P}$, independent
from $B_{t}$ and $K$ is the square integrable kernel%
\begin{equation*}
K_{H}\left( t,s\right) =C_{H}s^{\frac{1}{2}-H}\int_{s}^{t}(u-s)^{H-\frac{3}{2%
}}u^{H-\frac{1}{2}}\,du,\quad s<t
\end{equation*}%
($H>1/2)$. Then the process 
\begin{equation}
\eta _{t}^{\prime }=\exp \left( W_{t}-\frac{1}{2}t\right)   \label{2.11a}
\end{equation}%
is a square-integrable $\overline{P}-$martingale. Then, defining a standard
bi-dimensional Brownian motion,%
\begin{equation*}
W_{t}^{\ast }=(B_{t},W_{t})
\end{equation*}%
the process $\eta _{t}^{\ast }=\eta _{t}\eta _{t}^{^{\prime }}$%
\begin{equation*}
\eta _{t}^{\ast }=\exp \left\{ \int_{0}^{t}\Gamma _{s}\bullet dW_{t}^{\ast }-%
\frac{1}{2}\int_{0}^{t}\left\Vert \Gamma _{s}\right\Vert ^{2}ds\right\} 
\end{equation*}%
where, by lemma 2.1, $\Gamma =\left( \gamma ,1\right) $ satisfies the
Novikov condition, is also a $\overline{P}-$martingale. Then, by the
Cameron-Martin-Girsanov theorem, the process%
\begin{equation*}
\widetilde{W}_{t}^{\ast }=\left( \widetilde{W}_{t}^{\ast (1)},\widetilde{W}%
_{t}^{\ast (2)}\right) 
\end{equation*}%
defined by%
\begin{eqnarray*}
\widetilde{W}_{t}^{\ast (1)} &=&B_{t}-\int_{0}^{t}\gamma _{s}ds \\
\widetilde{W}_{t}^{\ast (2)} &=&W_{t}-t
\end{eqnarray*}%
is a bi-dimensional Brownian motion on the probability space $\left( 
\overline{\Omega },\overline{\mathcal{F}}_{T},Q_{T}^{\ast }\right) $, where $%
Q_{T}^{\ast }$ is the probability measure%
\begin{equation}
\frac{dQ_{T}^{\ast }}{d\overline{P}}=\eta _{T}^{\ast }=\eta _{T}\eta
_{T}^{^{\prime }}  \label{2.12a}
\end{equation}%
Moreover, the discounted price process $Z$ remains a martingale with respect
to the new measure $Q_{T}^{\ast }$. $Q_{T}^{\ast }$ being an equivalent
martingale measure distinct from $Q_{T}$, the market is incomplete.

\hfill $\blacksquare $

As stated above, incompleteness of the market is a reflection of the fact
that there are two different sources of risk and only one of the risks is
tradable. A choice of measure is how one fixes the volatility risk premium.

\section{Leverage and completeness}

The following nonlinear correlation of the returns 
\begin{equation}
L\left( \tau \right) =\left\langle \left\vert r\left( t+\tau \right)
\right\vert ^{2}r\left( t\right) \right\rangle -\left\langle \left\vert
r\left( t+\tau \right) \right\vert ^{2}\right\rangle \left\langle r\left(
t\right) \right\rangle   \label{C1}
\end{equation}%
is called \textit{leverage} and the \textit{leverage effect} is the fact
that, for $\tau >0$, $L\left( \tau \right) $ starts from a negative value
whose modulus decays to zero whereas for $\tau <0$ it has almost negligible
values. In the form of Eqs.(\ref{1.2a})(\ref{1.2b}) the volatility process $%
\sigma _{t}$ affects the log-price, but is not affected by it. Therefore, in
its simplest form the fractional volatility model contains no leverage
effect.

Leverage may, however, be implemented in the model in a simple way \cite%
{Vilela3} if one identifies the Brownian processes $B_{t}$ and $W_{t}$ in (%
\ref{1.2a}) and (\ref{2.10a}). Identifying the random generator of the
log-price process with the stochastic integrator of the volatility, at least
a part of the leverage effect is taken into account.

The identification of the two Brownian processes means that now, instead of
two, there is only one source of risk. Hence it is probable that in this
case completeness of the market might be achieved.

Identifying the random generators, a new fractional volatility model is
defined \cite{Vilela3}%
\begin{eqnarray}
dS_{t} &=&\mu _{t}S_{t}dt+\sigma _{t}S_{t}dW_{t}  \notag \\
\log \sigma _{t} &=&\beta +k^{\prime }\int_{-\infty }^{t}\left( t-u\right)
^{H-\frac{3}{2}}dW_{u}  \label{C2}
\end{eqnarray}%
\noindent \textbf{Proposition 3.1.} \textit{The market defined by (\ref{C2})
and (\ref{2.3}) is free of arbitrage and complete.}

\textit{Proof:} The proof of the first part of the proposition is analogous
to that of proposition 2.2. In fact, a similar argument to Lemma 2.2 yields
that%
\begin{equation*}
\eta _{t}=\exp \left\{ \int_{0}^{t}\frac{r-\mu _{s}}{\sigma _{s}}dW_{s}-%
\frac{1}{2}\int_{o}^{t}\left( \frac{r-\mu _{s}}{\sigma _{s}}\right)
^{2}ds\right\} 
\end{equation*}%
is a $P_{1}$-martingale with respect to $\left( \mathcal{F}_{1,t}\right)
_{0\leq t<T}$ and the probability measure $Q_{T}$, defined by $\frac{dQ_{T}}{%
dP_{1}}=\eta _{T}$ is an equivalent martingale measure.

Now that we have shown that the set of equivalent local martingale measures
for the market is non-empty, let $Q^{\ast }$ be an element in this set.
Then, recalling that $\left( \mathcal{F}_{1,t}\right) _{0\leq t<T}$ is the
augmentation of the natural filtration of the Brownian motion $W_{t}$, by
the Girsanov converse \cite{Musiela} \cite{Bjork} there is a $\left( 
\mathcal{F}_{1,t}\right) _{0\leq t<T}$ progressively measurable $\mathbb{R}$%
-valued process $\phi $ such that the Radon-Nikodym density of $Q^{\ast }$
with respect to $P_{1}$ equals%
\begin{equation*}
\frac{dQ^{\ast }}{dP_{1}}=\exp \left\{ \int_{0}^{T}\phi _{s}dW_{s}-\frac{1}{2%
}\int_{0}^{T}\phi _{s}^{2}ds\right\} 
\end{equation*}%
Moreover the process $W_{t}^{\ast }$ given by%
\begin{equation*}
W_{t}^{\ast }=W_{t}-\int_{0}^{t}\phi _{s}ds
\end{equation*}%
is a standard $Q^{\ast }-$Brownian motion and the discounted price process $Z
$ satisfies the following stochastic differential equation%
\begin{equation*}
dZ_{t}=(%
\mu
_{t}-r+\sigma _{t}\phi _{t})Z_{t}dt+\sigma _{t}Z_{t}dW_{t}^{\ast }
\end{equation*}%
Because $Z_{t}$ is a $Q^{\ast }-$martingale, then it must be hold $%
\mu
(t,\omega )-r+\sigma (t,\omega )\phi (t,\omega )=0$ almost everywhere w.r.t. 
$meas\times P$ in $\left[ 0,T\right] \times \Omega $, $meas$ being the
Lebesgue measure on the line. It implies%
\begin{equation*}
\phi (t,\omega )=\frac{r-%
\mu
(t,\omega )}{\sigma (t,\omega )}
\end{equation*}%
a. e. $\left( t,\omega \right) \in \left[ 0,T\right] \times \Omega _{1}$.
Hence $Q^{\ast }=Q$, that is, $Q$ is the unique equivalent martingale
measure. This market model is complete.

\hfill $\blacksquare $

\section{Remarks and conclusions}

1) Partially reconstructed from empirical data, the fractional volatility
model describes well the statistics of returns. The fact that, once the
parameters are adjusted by the data for a particular observation time scale $%
\delta $, it describes well different time lags is related to the fact that
the volatility is driven not by fractional Brownian motion but its
increments.

Specific trader strategies and psychology should play a role on market
crisis and bubbles. However, the fact that in the fractional volatility
model the same set of parameters would describe very different markets \cite%
{Oliveira} seems to imply that the market statistical behavior (in normal
days) is more influenced by the nature of the financial institutions (the
double auction process) than by the traders strategies \cite{Vilela2}.
Therefore some kind of universality of the statistical behavior of the bulk
data throughout different markets would not be surprising.

The identification of the Brownian process of the log-price with the one
that generates the fractional noise driving the volatility, introduces an
asymmetric coupling between $\sigma _{t}$ and $S_{t}$ that is also exhibited
by the market data.

2) In this paper, mathematical consistency of, both versions, of the
fractional volatility model has been established. This and its better
consistency with the experimental data, makes it a candidate to replace
geometrical Brownian as the standard market model.

\end{document}